\documentclass[10pt]{article}
\usepackage[tmargin=1in,bmargin=1in,lmargin=1.25in,rmargin=1.25in]{geometry}

\usepackage{setspace}
\usepackage{csquotes}
\usepackage{authblk}
\usepackage{amsmath}
\usepackage{amsfonts}
\usepackage{amsthm}
\usepackage{stmaryrd}
\usepackage{mathtools}
\usepackage{graphicx}
\usepackage{comment}
\usepackage[title]{appendix}
\usepackage[font=small,labelfont=bf]{caption}
\usepackage{mwe}
\usepackage{libertine}
\usepackage[libertine]{newtxmath}
\usepackage{microtype}
\usepackage{anyfontsize}
\usepackage{footmisc}

\usepackage[natbib,style=authoryear-icomp,backref=true,backend=biber]{biblatex}
\usepackage{xcolor}
\definecolor{myurlcolor}{rgb}{0,0,0.4}
\definecolor{mycitecolor}{rgb}{0,0.5,0}
\definecolor{myrefcolor}{rgb}{0.5,0,0}
\usepackage[breaklinks=true]{hyperref}
\hypersetup{colorlinks,
linkcolor=myrefcolor,
citecolor=mycitecolor,
urlcolor=myurlcolor}
\usepackage[capitalize]{cleveref}

\newcommand{\R}{\mathbb{R}}
\newcommand{\N}{\mathbb{N}}

\theoremstyle{definition}
\newtheorem{definition}{Definition}[section]

\theoremstyle{plain}
\newtheorem{theorem}[definition]{Theorem}

\usepackage{todonotes}

\title{Hyperdeterminism? Spacetime `Analyzed'}
\author{Lu Chen\thanks{University of Southern California. Email: chen.l@usc.edu}, \, Tobias Fritz\thanks{University of Innsbruck. Email: tobias.fritz@uibk.ac.at} 
}

\begin{document}
	\maketitle

\begin{abstract}

When modelling spacetime and classical physical fields, one typically assumes smoothness (infinite differentiability). But this assumption and its philosophical implications have not been sufficiently scrutinized. For example, we can appeal to analytic functions instead, which are also often used by physicists. Doing so leads to very different philosophical interpretations of a theory. For instance, our world would be `hyperdeterministic' with analytic functions, in the sense that every field configuration is uniquely determined by its restriction to an arbitrarily small region. Relatedly, the hole argument of general relativity does not get off the ground. We argue that such an appeal to analytic functions is technically feasible and, conceptually, not obviously objectionable. The moral is to warn against rushing to draw philosophical conclusions from physical theories, given their drastic sensitivity to mathematical formalisms.

\end{abstract}

\section{Introduction}

In classical field theory---including general relativity---one considers field configurations subject to field equations.
There is a sophisticated mathematical theory of the existence and uniqueness of solutions to these equations for given initial conditions or boundary conditions.
The mathematical theorems involved make different assumptions about the regularity of the field configurations under consideration, such as the degree of differentiability.
In the physics literature, typically either smoothness (differentiability to all orders) or analyticity (existence of Taylor expansions) is assumed.
But these assumptions have never been properly examined or compared on philosophical grounds.
We aim to remedy this in the present paper.

The fact that this topic has been largely ignored is most evident when we turn to the hole argument,
which purports to show that the world is indeterministic under general relativity.
Although the hole argument has been much discussed, its smoothness assumption on the metrics and diffeomorphisms in spacetime models has never been scrutinized.\footnote{Perhaps the most prominent discussion of analytic functions in the context of general relativity occurs in the literature of black hole uniqueness theorems (in the vicinity of the singularity theorems) which often assume analyticity. However, the discussion centers around the removal of the analyticity assumption rather than any serious evaluation of the assumption (see \Cref{analytic_gr_fn}).} We observe that the hole argument would not go through if we resort to analytic field configurations (and accordingly, analytic diffeomorphisms).
If we appeal to analytic functions instead, we would be led to the opposite conclusion that the world is `hyperdeterministic': the physical state on an arbitrarily small open region of spacetime is sufficient for determining the physical state on the whole spacetime.

Using smooth functions is a typical working assumption of physicists in field theory and general relativity, but this is not forced upon us by mere technical considerations. Indeed standard differential geometry and general relativity make sense just the same with other classes of functions, as long as the technical requirement that the module of derivations is finitely generated projective is satisfied~\citep{fritz}.


Our plan is as follow. 
First, we introduce and motivate analytic functions based on their widespread use in physics (\Cref{sec_anal}).
We then explain how we can formulate classical field theories in terms of analytic functions in \Cref{analytic_physics}, and turn to analytic spacetime theories in \Cref{analytic_gr}, showing how they are technically and empirically adequate.
We then extract philosophical consequences, including the blocking of the hole argument and hyperdeterminism.
Finally in \Cref{hyperdet_phil}, we evaluate hyperdeterminism on philosophical grounds.

\section{Preliminaries on analytic functions}
\label{sec_anal}

Like smooth functions, analytic functions are infinitely differentiable, but they are yet more restrictive:
	
	\begin{definition}
		A function $f: U \rightarrow \R$ defined on an open set $U \subseteq \R^n$ is \emph{analytic}\footnote{The proper mathematical term would be \emph{real-analytic} in order to distinguish these functions from \emph{complex-analytic} (holomorphic) functions, but we consider real-analytic functions only and omit the `real' in order to keep the terminology as simple as possible.} if it is infinitely differentiable and for every point $a$ in $U$, there exists a power series
		\begin{equation}
			\label{power_series}
			\sum_{j_1, \ldots, j_n = 0}^\infty c_{j_1, \ldots, j_n} (x_1 - a_1)^{j_1} \cdots (x_n - a_n)^{j_n}
		\end{equation}
		that converges to $f(x)$ for every $x$ in some neighborhood of $a$ inside $U$ (where $x_i$ and $a_i$ are the components of $x$ and $a$). 	
	\end{definition}
 If we consider just a one-dimensional space, then the power series \eqref{power_series} simplifies to $\sum_{j=0}^{\infty}c_j(x-a)^j$. We can see that polynomial functions such as $f(x)=ax^2+bx+c$ are trivially analytic. Also, the exponential function $x \mapsto e^x$ is expandable into such power series, and is therefore an analytic function on $\R$.\footnote{Namely it is expandable into the power series $\sum_{j=0}^{\infty} \frac{e^a}{j!} (x - a)^j$ around any $a \in \R$.} For any analytic function, its corresponding power series is necessarily given by the \emph{Taylor expansion} of $f$ around $a$, which can be obtained by taking partial derivatives in the equality between $f(x)$ and~\eqref{power_series}  and plugging in $x = a$.\footnote{More explicitly, the coefficients can be written as
\[
	c_{j_1, \dots, j_n} = \frac{1}{j_1! \dots j_n!} \cdot \frac{\partial^{j_1 + \dots + j_n} f}{\partial x_1^{j_1} \dots \partial x_n^{j_n}}(a)
	\qquad \forall j_1, \ldots, j_n \in \N.
\]}

The striking feature of analytic functions central to our ensuing discussion can be captured by the `identity theorem': if two analytic functions agree on an arbitrarily small open set, then they agree everywhere on the domain of definition:

\begin{theorem}[Identity Theorem]
	\label{identity_theorem}
	Let $f$ and $g$ be analytic functions on a connected open set $U \subseteq \R^n$.
	If $f$ and $g$ coincide on a nonempty open subset of $U$, then they coincide on all of $U$.
\end{theorem}

More generally, a function $f : U \to \R^m$ for arbitrary $m \in \N$ is called analytic if all of its component functions $f_1, \dots, f_m : U \to \R$ are, and the identity theorem immediately generalizes to such functions.

Before we turn to a more detailed discussion of the scope of analytic functions in physics, let us first underline their widespread use in physics.
First, when working non-rigorously, physicists often make use of Taylor expansions without worrying about the series' convergence.
This amounts to an implicit assumption of `analyticity by default' on the functions involved.\footnote{This automatic assumption will be questioned even by non-rigorous physicists if it turns out to lead to inconsistencies further down the line. This is why we call it `analyticity by default'.}
In rigorous settings, analytic functions also appear frequently, such as the correlation functions in quantum field theory (in the form of the Schwinger functions), and in many different ways in mathematical relativity\footnote{\label{analytic_gr_fn} For example, versions of black hole uniqueness theorems tend to assume analyticity~\citep{heusler}, and maximal analytic extensions of spacetimes are an important topic~\citep{BL}. 
On the other hand, eliminating analyticity assumptions from theorems of mathematical relativity is the goal of some contemporary research~\citep{AIK,CG}.}, where this assumption is often regarded as mathematically convenient though perhaps otherwise undesirable.
As we will see below, solutions to ordinary differential equations (like equations of motion in classical mechanics) and partial differential equations (like field equations) tend to be analytic. For example, the electric field associated to a static and analytic distribution of charge is analytic as well.
Furthermore, analytic functions are dense in the Hilbert space of wave functions $L^2(\R^3)$,\footnote{For example, the eigenfunctions of the quantum harmonic oscillator are analytic, and their finite linear combinations are still analytic and dense in $L^2(\R^3)$.}
which means that for many purposes, wave functions can be assumed analytic without loss of generality.

\section{Analytic field theories}
\label{analytic_physics}

How exactly do we do physics in terms of analytic functions? It is instructive to first consider classical field theories without general covariance, where already a number of interesting features of physics with analytic functions become apparent.
We will then turn to general relativity and the hole argument in the next section.

A classical field theory is described by a field equation for a field configuration\footnote{It is enough for our purposes to consider the case where $\phi$ takes values in $\R^d$ for some $d$, since the points that we make all apply similarly to the case where $\phi$ is a section of a fiber bundle (as e.g.~in gauge theory).} $\phi$, which is typically a set of partial differential equations (PDE) of the form\footnote{Analogous statements hold true 
also for theories without time, like electrostatics. Their field equations are elliptic PDEs, which tend to have analytic solutions as well~\citep{morrey}. There also exist similar results for theories involving higher time derivatives. 
}
\begin{equation}
	\label{field_equation_general}
	\partial^2_t \phi = F(\phi, \partial_i \phi, \partial_t \phi, \partial_i \partial_j \phi, \partial_i \partial_t \phi, x), \qquad
	\phi(x, 0) = f(x), \qquad \partial_t \phi(x, 0) = g(x),
\end{equation}
where $i$ and $j$ are indices ranging over spatial directions, $F$ is itself analytic and the initial conditions $f$ and $g$ are analytic as well.
For PDEs of this type, the classical Cauchy--Kovalevskaya theorem~\citep[Seection~1.D]{folland} shows that a unique solution\footnote{The existence of a solution means \emph{local} existence, as it is often the case that the this solution `blows up' in finite time, so that a solution valid for all times does not exist. This may signal the limit of applicability of the physical theory under consideration. For an interesting example where it is a famous open problem whether blow-up happens (with smooth initial conditions), consider the Navier--Stokes equations of fluid mechanics~\citep{constantin}.} exists, and this solution is guaranteed to be analytic as well.
A simple special case where this applies is systems in classical mechanics, where $\phi$ is the list of coordinates of all particles in the system and $x$ has zero components, i.e., $\phi$ is a function of only the time $t$.

Can we infer from this that the physics of such a system can be adequately modelled by analytic functions?
By the Cauchy--Kovalevskaya theorem, this amounts to asking whether the right-hand side of the field equation $F$ and the initial conditions $f$ and $g$ can be expected to be analytic.
For the right-hand side $F$, this is usually the case, in particular for those PDEs that formalize basic physical laws.
On the other hand, when modelling systems containing solid materials, physicists also frequently consider systems where $F$ is not even smooth; a good example would be the Maxwell equations in a medium with a discontinuous dielectric constant (due to two different materials interfacing).\footnote{Another interesting but less physical example is Norton's dome, where the lack of uniqueness of a solution to the equation of motion---and hence the apparent lack of determinism---can be attributed to the fact that Norton's $F$ is not analytic (and once again not smooth either), so that the Cauchy--Kovalevskaya theorem does not apply.}
Similar statements apply to the initial conditions $f$ and $g$: they are often analytic, but sometimes one is interested in situations where they are not even smooth.
In fact, we know of no realistic physical model in which either one of $F$, $f$ or $g$ would be smooth but not analytic.

So functions that are merely $k$-fold continuously differentiable for some $k \in \N$ do play a role in physics, which may suggest that the analytic-function framework is too restrictive. 
But such occurrences of nonanalytic functions in effective theories do not require their occurrence in more fundamental theories. When we model the whole universe, a physical theory need not even require initial conditions (e.g., the FLRW model; see Section 4), and in this case the question of their analyticity does not even arise.
Furthermore, note that restricting models with initial conditions to analytic functions is almost guaranteed to be empirically adequate.
This is because  initial conditions, construed as $k$-fold continuously differentiable functions, can be approximated by analytic functions arbitrarily well.\footnote{More precisely, the analytic functions are dense in the Sobolev space $W^{k,p}(\R^n)$ for every $k \in \N$ and $p \in [1,\infty)$. We show that this is the case in three steps. We first appeal to the \emph{Meyers--Serrin theorem}~\citep[Theorem~11.24]{leoni}, which establishes the density of smooth functions in $W^{k,p}(\R^n)$. Second, compactly supported smooth functions are dense in all smooth functions. Third, a compactly supported smooth function can be approximated in Sobolev norm by the sequence of analytic functions obtained by convoluting it with Gaussians of smaller and smaller width.}
For many of the usual PDEs in physics, such an approximation of initial conditions results in a good approximation to the whole history, since field equations tend to be such that a small perturbation to the initial conditions $f$ and $g$ translates into a small perturbation of the solution.\footnote{Formally, the Cauchy problem for those PDEs that occur in physics is \emph{well-posed}~\citep[Section~1.5]{strauss}.}


It is worth noting again that a proponent of smooth functions faces similar concerns, given that e.g.~the electromagnetic field in a medium with a discontinuous dielectric constant is not even going to be smooth.
Even worse, smooth functions do \emph{not} satisfy the analogue of the Cauchy--Kovalevskaya theorem, as there are examples of PDEs as in~\eqref{field_equation_general} with smooth $F$ and without \emph{any} smooth solution for any initial conditions~\citep{lewy}.\footnote{It is still plausible that the PDEs that are relevant for physics and in which $F$ is smooth have smooth solutions, so we do not view this as a strong argument against smooth functions.}

To take stock, we do not see any technical or empirical reasons against using analytic functions in classical field theories.
Nevertheless, one may argue that there are still no good reason to think that field configurations \emph{really} are analytic.
Indeed one may argue that restricting to analytic functions is objectionable qua restriction, and granting that analytic functions are widely used in physics, they may be mere idealizations made for practical convenience.\footnote{As we have emphasized, using smooth functions also amounts to a restriction from a wider class of functions. But here, the concern is that this is a problem for smooth functions too, and the case of analytic functions is more problematic still given that they are more restrictive than smooth functions. We thank the anonymous referee for pressing on this.}
To respond, we take all technically and empirically adequate mathematical models seriously, considering their constituents to be serious candidates for realist commitment (see for example~\cite{wallace}).
Furthermore, we do not consider a restriction of modelling options a drawback per se.
As an analogy, \citet{Belot} proposed a relationist revision of Newtonian mechanics, which entails that the total angular momentum vanishes.
This is not considered a flaw of the theory, but rather a virtue~\citep{Pooley2}.

\section{Analytic general relativity and the hole argument}
\label{analytic_gr}

Let us now turn to general relativity in terms of analytic functions.
Call it `analytic general relativity'.
Similar to the previous section, this amounts to restricting all physical fields, first and foremost the metric field, to be analytic.
Secondly, diffeomorphism symmetry must likewise be restricted to `analytic maps', because the pushforward or pullback of an analytic field by a diffeomorphism should preserve its analyticity. Lastly, a manifold is as how it behaves, and thus we need to resort to `analytic manifolds'.\footnote{Diffeomorphisms are maps that preserve all structure of manifolds. Manifolds whose structures are preserved by analytic diffeomorphisms only must be ipso facto analytic manifolds.} 
While nothing in the following is technically striking, we shall give an explicit presentation of the framework as we are not aware of existent ones in the literature.

Let us formalize these concepts in the reverse order (that is, the order of definitions reverses the order of justification). First, analytic manifolds are manifolds that can be locally mapped to $\R^n$ by analytic functions. More technically: 

\begin{definition}[Analytic Manifold]
	An \textit{analytic manifold} of dimension $n$ is a topological manifold equipped with an atlas where:
	\begin{enumerate}
		\item Each chart is a homeomorphism from an open subset of the manifold to an open subset of $\R^n$.
		\item For any two overlapping charts, the transition map (the map from an open subset of $\R^n$ to another via the manifold) is required to be an analytic function.
			In other words, in each overlap the coordinate change is described by functions that can be written around each point as a convergent power series.
	\end{enumerate}
\end{definition}

Then regarding diffeomorphisms, a map $f : M \to N$ between two $n$-dimensional analytic manifolds is an \textit{analytic diffeomorphism} if its local expression in charts is analytic\footnote{Meaning that if $\phi : U \to M$ and $\psi : V \to N$ for $U, V \subseteq \R^n$ are charts, then $\psi^{-1} \circ f \circ \phi$ should be analytic on its domain of definition.}, and if it is bijective with a likewise analytic inverse $f^{-1} : N \to M$.

Regarding fields, an \textit{analytic tensor field} of type $(k, l)$ on an analytic manifold $M$ is a map that assigns to each point $p \in M$ a tensor $T_p$ of type $(k, l)$ in the tensor product of the tangent and cotangent spaces at $p$, namely $(T_pM)^{\otimes k} \otimes (T_p^*M)^{\otimes l}$. Furthermore, in any coordinate chart with coordinates $\{x^i\}$, the components $T_{j_1 \ldots j_l}^{i_1 \ldots i_k}(x)$
of the tensor field are required to be analytic functions of the coordinates $x^1, \ldots, x^n$.
In particular, we can speak of the metric field $g_{\mu\nu}$, which is a (0,2) tensor field, being analytic.

	Consider now a $4$-dimensional analytic manifold $M$ representing spacetime. The Einstein Field Equations (EFEs) in terms of analytic fields take the usual form
	\begin{equation}
		\label{efe}
		G_{\mu\nu} + \Lambda g_{\mu\nu} = \kappa T_{\mu\nu},
	\end{equation}
	which is now considered a field equation for an \emph{analytic} metric field $g$, which implies that the Einstein tensor $G_{\mu\nu}$ is analytic as well\footnote{Because it is a polynomial expression of the metric, its inverse and its first and second derivatives.}, and the equation itself shows that the stress-energy tensor $T_{\mu\nu}$ must be analytic too.
	 
	We now say that a solution $g$ of the EFEs is \emph{analytic} if it is an analytic metric field on an analytic manifold $M$ and the EFEs hold.\footnote{This is not to be confused with the informal term of an `analytic solution', often used to refer to solutions that can be explicitly written down in closed form rather than merely numerically.}
	Many commonly seen solutions $g$ of the EFEs are analytic in nature: there is an obvious analytic structure on the underlying manifold that makes them analytic. For example, flat Minkowski spacetime is one of the vacuum ($T_{\mu\nu} = 0$) solutions, and it is clear that this metric field is analytic\footnote{With respect to the obvious analytic structure on $\R^4$ given by the identity as the only chart.}, since at each spacetime point the metric is represented by the diagonal matrix $\mathrm{diag}(-1,1,1,1)$, whose entries are trivially analytic.\footnote{If we had chosen a suitably \emph{different} analytic manifold structure on $\R^4$---one which is related by the usual one via a non-analytic smooth diffeomorphism---then the Minkowski metric would \emph{not} necessarily be analytic. So the point that we make here is that \emph{there is} an analytic structure on the manifold which makes the metric analytic.} 

	 As a less trivial example, the Friedmann-Lemaître-Robertson-Walker (FLRW) metric, which is central to modern cosmology and describes a homogeneous, isotropic, path-connected, expanding universe, is given by:
	 \begin{equation}
	 	ds^2 = -c^2dt^2 + a(t)^2\left[\frac{dr^2}{1 - kr^2} + r^2(d\theta^2 + \sin^2\theta d\phi^2)\right].
	 \end{equation}
	 Here, $a(t)$ is the scale factor modeling the expansion of the universe as a function of time, $k$ is a parameter determining the curvature of space, and $(t, r, \theta, \phi)$ are the comoving coordinates.\footnote{\emph{Comoving} means that the lines of constant $r$, $\theta$ and $\phi$ are geodesics describing the worldlines of galaxies that are at rest relative to the cosmic microwave background.}
	 The EFEs are relevant only for determining $a(t)$, while the form of the metric itself results from the assumption of isotropy and homogeneity. Whether the FLRW metric is analytic or not---in the analytic structure determined by the comoving coordinates---depends on whether $a(t)$ is an analytic function of $t$.
	 With the big bang singularity at $t=0$ excluded as usual, this function is indeed typically analytic.\footnote{To view some options, arising for various choices of matter content determining the form of $T_{\mu\nu}$:
	 	\begin{itemize}
	 		\item Matter-Dominated Universe: $a(t) \propto t^{2/3}$
	 		\item Radiation-Dominated Universe:$a(t) \propto t^{1/2}$
			\item Dark Energy-Dominated Universe (Cosmological Constant): $a(t) \propto \exp\left(c t \sqrt{\Lambda/3} \right)$
	 		\item Inflationary Models: $a(t) \propto e^{Ht}$, where $H$ is the Hubble parameter during inflation.
	 	\end{itemize}
	 	All the functions on the right-hand sides of these equations are analytic functions of $t>0$.}
	 
	 Or consider the Schwarzschild metric, another vacuum solution of the EFEs that describes the gravitational field around a spherical mass that has no electric charge or angular momentum, such as a non-rotating black hole. 
The metric, given as follows, is analytic within its domain of definition in the usual Schwarzschild coordinates, namely
	 	\begin{equation}
	 		ds^2 = -\left(1 - \frac{2GM}{c^2r}\right)c^2dt^2 + \left(1 - \frac{2GM}{c^2r}\right)^{-1}dr^2 + r^2(d\theta^2 + \sin^2\theta d\phi^2)
	 	\end{equation}
		where as usual the black hole singularity at $r = 0$ is not part of the spacetime manifold, and the coordinate singularity at the event horizon located at the Schwarzschild radius $r_s = \frac{2GM}{c^2}$ is excluded from the chart.

		In fact, the analyticity of the Schwarzschild solution is no singular occurrence: 
		a theorem due to~\citet{mzh} states that \emph{every} stationary vacuum solution of the EFEs is analytic, and this applies to the Schwarzschild metric in particular. Similarly, \citet[Theorem~1]{BS} have shown that a stationary asymptotically flat spacetime is automatically analytic at spatial infinity.
		Results like these suggest that the analyticity of solutions to the EFEs is not a rare exception, but rather a common feature of general relativity.

	Nevertheless, it is worth clarifying that analytic solutions are obviously not \textit{comprehensive}: we can easily come up with spacetime models that are not analytic. For example, we can simply change the analytic structure on the underlying smooth manifold by writing the metric in any of the above examples in coordinates that differ from the above ones by a suitable non-analytic coordinate transformation.
	 More drastically, a metric that is flat within a region but has nonzero scalar curvature elsewhere is not analytic for \emph{any} analytic structure on the underlying manifold; this is a straightforward consequence of the identity theorem.

	 But again, it is worth emphasizing that all of this is just as it is for smooth functions.
	 Like the electromagnetic case in the previous section, neither analytic nor smooth fields are sufficiently broad to encompass all physical models in general relativity. For instance, the mass distribution and the metric of a relativistic star are not smooth~\citep[Box~23.2]{MTW}, since the mass density drops to zero non-smoothly at the surface of the star. 
	 There are other models of physical interest, such as cosmic strings, where even the concept of function itself needs to be generalized~\citep{SV}.
	 But this does not mean that smooth (or analytic) functions are empirically inadequate, since one can reasonably expect that the mass distribution and the metric can be approximated by such functions while preserving the EFEs.\footnote{For rigorous results along these lines, see the literature on stability in general relativity, such as~\citep{CD}. It is worth noting that besides such highly technical stability results, also examples of instability in general relativity are known, where small perturbations in initial conditions lead to large perturbations in the global solution~\citep{moschidis}.}

\subsection{The hole argument}

As we mentioned in the beginning of the paper, if we appeal to analytic functions as opposed to smooth ones, then the hole argument would not go through.
Instead we run into an opposite `trouble': hyperdeterminism. To remind: our world is \textit{hyperdeterministic} if the physical state on an arbitrarily small open region of spacetime is sufficient to determine the physical state on the whole spacetime. We can appeal to possible worlds to unpack this further: for any physically possible world $w$ that is identical to the actual world on an arbitrarily small open region of spacetime, $w$ is identical to the actual world.
This result is straightforward given the identity theorem, but let's make it explicit.

To start, let's recall the hole argument. A spacetime model is written as $\langle M, g, \Psi\rangle$, where $M$ is a connected smooth manifold, $g$ the metric field, and `$\Psi$' the collection of all other matter fields. Consider an open subregion $\mathcal{H}$ (the `hole') properly contained in $M$. A hole transformation is a diffeomorphism $f : M \to M$ that is nonidentity on $\mathcal{H}$ and identity outside of it. According to the principle of general covariance, if $\langle M, g, \Psi\rangle$ is a solution to the EFEs, then so is $\langle M, f_*g, f_*\Psi\rangle$.
The models before and after the hole transformation are regarded as naively distinct, although they are empirically indistinguishable because the physical content of general relativity (or whichever generally covariant theory one considers) is invariant under diffeomorphisms.
This leads to the  problem of indeterminism: even if we know the whole world history outside of the hole, we would not be able to 
uniquely determine the field configuration inside it, no matter how small the hole is. This is---as the argument goes---presumably problematic because indeterminism occurs `too easily'.

The hole argument does not go through in analytic general relativity because the hole transformation is not possible with analytic diffeomorphisms.
That is, for any generic open subregion of a connected spacetime manifold $M$, there are no analytic diffeomorphisms that are the identity map outside the region and non-identity inside it.\footnote{Formally, if $\mathcal{H} \subseteq M$ is open such that $M \setminus \mathcal{H}$ has nontrivial interior, then any diffeomorphism $f : M \to M$ that is identity on $M \setminus \mathcal{H}$ must be the identity. Indeed if we write $V \subseteq \mathcal{H}$ for the nonempty open set on which $f$ differs from the identity, then by connectedness of $M$ we have $\overline{V} \cap (M \setminus V) \neq \emptyset$. Choosing a chart around any $x \in \overline{V} \cap (M \setminus V)$ and expressing $f$ in coordinates with respect to this chart then contradicts \Cref{identity_theorem}, since the chart overlaps both with $V$ and with the interior of $M \setminus V$.}
Thus, for a connected analytic spacetime model $\langle M, g, \Psi\rangle$ that solves the EFEs exactly, there is no distinct diffeomorphically related analytic model $\langle M,f_*g,f_*\Psi\rangle$ that is also a solution.
In fact, we are now led to the other extreme: given any open subregion of $M$, no matter how small, there are no two spacetime models that agree on this region but disagree elsewhere.
Thus hyperdeterminism. 

To take stock, while smooth functions seem to result in over-flexibility in transformations that result in equally physically adequate models, analytic functions seem to result in over-rigidity, where global spacetime content is determined by local properties.

In the face of this dilemma, it is natural to ask whether there is a way out.
For example, could we resort to constructions based on a different class of functions that hit the sweet spot in the middle, neither too rigid nor too flexible? The answer, however, seems to be negative given the principle of general covariance featured by general relativity\footnote{Without general covariance, one could for example try to consider a notion of function that is analytic in the temporal direction but merely smooth in the spatial directions. Using such functions as field configurations could result in the intended rigidity in the time evolution while maintaining flexibility with regards to spatial initial conditions.}:
if the identity principle holds for a given class of functions, then hyperdeterminism obtains; whereas if the identity principle fails, then its failure suggests that for some connected manifold $M$, there is a non-identity diffeomorphism $f : M \to M$ that is identity on some nonempty open set $V \subseteq M$\footnote{Note that making this argument fully precise would first of all require us to be more precise about what a general `class of functions' is and how these functions determine the class of allowed diffeomorphisms, an undertaking which we will not pursue here. But roughly, if functions $g, h : U \to \R$ violate the identity theorem, then this means that they coincide on a nonempty open set $V \subseteq U$ but differ outside of it. It is then natural to consider the function $\tilde{f} : U \to \R^n$ defined by $(x_1, \dots, x_n) \mapsto (g(x) - h(x) + x_1, x_2, \dots, x_n)$. This function is the identity on $V$ but differs outside of it. The remaining problem (which we do not address here) is to prove that $g$ and $h$ can be chosen such that $\tilde{f}$ diffeomorphically maps some open set $M \subseteq U$ containing $V$ to itself, so that one can take $f \coloneqq \tilde{f}|_M$.}, and the hole argument unfolds with $\mathcal{H} \coloneqq M \setminus \overline{V}$.\footnote{Since $f$ is the identity on $V$, the same is true on its closure $\overline{V} = M \setminus \mathcal{H}$.}


\section{Hyperdeterminism}
\label{hyperdet_phil}

As we already noted, due to the identity theorem, (classical) physics with analytic functions results in hyperdeterminism: a field configuration is uniquely determined by what it looks like on an arbitrarily small open region.
While this is surely counterintuitive and at odds with how we typically think of the flexibility in field configurations and initial conditions, the empirical adequacy we already mentioned indicates that there may not be anything inherently problematic about it.\footnote{This is suggested already by the common appeal to maximal analytic extensions in general relativity, where the analyticity assumption on the metric and manifold are not put into question despite being crucial for the uniqueness of the extension~\citep{BL,CC}.}
Let's discuss these issues in more detail now.
Note that we do not claim that there is no conceptual objection to hyperdeterminism (which would be outside the scope and aim of this paper), but we provide a quick survey of all obvious objections and explain how they are not severe.


What might be objectionable about hyperdeterminism? Perhaps one may object to it based on how radically different it is from the intuitive idea behind determinism, namely the idea of the past `generating' or `evolving into' the future under the laws of nature. Clearly hyperdeterminism is not amiable to housing such an intuition---we certainly do not entertain an analogous idea of the states in our lab `generating' or `evolving into' those in your lab at space-like separation. However, while some philosophers---such as~\citet{maudlin}---do take the intuitive idea of determinism seriously, many have pointed out that this idea is in tension with our existing physical theories~\citep{CG2,adlam}. For example, our spacetime may fail the global hyperbolicity that is required for the concept of time slice to be well-defined. A generalized definition of determinism that is not based on a temporal direction can appeal to a generic spacetime region as `the generator'. For example, \citet{butterfield} proposed that for a kind of region $S$, a theory is called `$S$-deterministic' if two spacetime models of the theory that involve $S$ agree everywhere if they agree on $S$.\footnote{We have simplified his proposal, because the difference, while important to his discussions, is irrelevant to ours.} Here, $S$ is not necessarily a time slice. Hyperdeterminism is precisely a version of  $S$-determinism, where $S$ is understood as any open spacetime region. 
The more recent definitions of determinism proposed by \citet{adlam}, which are formulated in a constraint-based framework of physical laws, also do not discriminate against hyperdeterminism as a legitimate instance of determinism. Moreover, if one adopts strong determinism  about our world---according to which there is only one possible physical history~\citep{penrose,chen}---then hyperdeterminism is just as reasonable as the naive `evolving into' determinism.



One may counter that the determinism displayed by the usual laws of (classical) physics is \emph{dynamical} in nature, as it is a consequence of the equations of motion or field equations, while the hyperdeterminism associated with analytic function physics is \emph{kinematical} in the sense that it already obtains even without imposing dynamical equations.
But it is not clear why determinism \textit{should} be dynamical in nature. If the worry is that we would like to retain the possibility of non-deterministic dynamics, then we can also appeal to a different function space than the analytic one in that case. 
Moreover, this attempted counterargument runs into the difficulty that many physical theories display no clear boundary between kinematical and dynamical conditions.
For example in electrodynamics in the vacuum, Gauss' law $\nabla \vec{E} = 0$ is dynamical as long as one considers the electric and magnetic fields $\vec{E}$ and $\vec{B}$ as the real physical fields, while it becomes purely kinematical if one considers the four-potential $A^\mu$ as the physically real field (which is empirically equivalent).\footnote{This blurry boundary is more discussed in the context of special relativity for example in~\citet{pooley,CR}. As a case in point, the Minkowski metric $\eta_{\mu\nu}$ can be seen as kinematically fixed in all models, or can be dynamically fixed by imposing the additional `dynamical law' $R^\rho_{\phantom{\rho}\gamma\mu\nu}=0$.}

A second objection might be that hyperdeterminism violates the Humean--Lewisian principle of free recombination.\footnote{Although it is unclear to what extent we should stick to a priori metaphysical principles contra other considerations, let us entertain the idea that physical theories should abide by this principle and derive the consequences of this.}
	This principle postulates that any distribution of perfectly natural properties is possible, and is adapted from the Humean principle that there are no necessary connections between distinct existences~\citep{lewis}.
	On first look, it seems that this principle is violated in analytic physics because field values in two disjoint regions can be incompatible, assuming that field values at a spatial region are perfectly natural properties of that region (see the footnote for  more nuance).\footnote{
Strictly speaking, hyperdeterminists need not claim that hyperdeterminism holds in every possible world (we do not need to assume that every metaphysically possible world is modeled by analytic functions) and therefore need not be at odds with this principle. But given hyperdeterminism, certain combinations of natural properties would not be as close a possibility as one might think and therefore our modal intuition is still violated to an extent. For example, we might think that a world that is the same as our world except that it has regions with zero values, the counterpart of which have near-zero values in our world, is a close possibility. But given hyperdeterminism, it would be a more remote possibility in which hyperdeterminism is not true.}
But then again, already continuous functions seem to violate---though minimally---this principle: the value at one spatial point already imposes a constraint on the values at neighboring points (that is, they need to converge to this value at this point).
A similar kind of tension between free recombination and physics appears in the form of constraints.\footnote{There is an interesting parallel discussion in~\citep{vassallo} on the implication of the Bianchi identities in general relativity for the Humean--Lewisian metaphysics. The idea is that the Bianchi identities, which are necessary mathematical constraints on the Einstein tensor, is a physically significant part of the laws of general relativity. This raises the question whether laws of nature can be necessary in the strong Kripkean sense, which challenges the Humean view that laws are contingent. This parallels the apparent challenge of our analytic formalism to the Humean principle of free recombination. While these discussions are concerned with different aspects of the Humean--Lewisian metaphysics,  both have to do with necessary connections (one concerns necessary connections between entities whose properties are involved in laws and the other, of course, is about necessary connections between properties in disjoint spacetime regions).} Recall that a constraint is a nontrivial condition that a spatial field configuration must satisfy in order to qualify as a physical initial condition.
	Consider again Gauss's law $\nabla \vec{E} = 0$ (assuming the absence of charges for simplicity), which does not involve any time derivatives and therefore imposes a nontrivial condition that a vector field on $\R^3$ must satisfy in order for it to qualify as a possible configuration of the electric field at any given time.\footnote{But as we already noted, this constraint can be eliminated by considering the four-potential as the physical field instead.}

	Nonetheless, it is reasonable to expect \textit{some} version of the principle of free recombination to hold, since we may still want---for example---counterfactual explanations in physics.  For instance, to show the wave-like nature of an electron, we can appeal to the facts that (1) an electron shot through a double slit barrier hit a screen behind somewhere in the middle; (2) if it were shot through the barrier where one slit is covered, it would not have hit such a spot (but instead somewhere directly behind the open slit). If hyperdeterminism is true, how do we make sense of counterfactual statements like (2), given that we fix the rest of the experimental setup? Clearly, assuming hyperdeterminism, such a statement would be \textit{vacuously} true if we strictly fix the properties of any spacetime region, which would be undesirable. To respond, we consider such counterfactual statements to be made true by models of counterfactual physical situations that are solutions to relevant dynamical equations. These different models not only differ within the spacetime region consisting of the experiment itself, but also \emph{outside} of where the experiment happens: for example, one of them features a plan for closing one slit while the other does not---indeed, there is no region on which these models  are strictly identical.\footnote{One may argue that this is counterintuitive: in order to make a counterfactual true, we need to slightly distort the whole spacetime `mosaic'. But is this account really implausible? While we do want to keep certain features of the  world invariant in counterfactual reasonings, this invariance needs only to be approximate. At least when physicists make counterfactual statements, we observe no requirement on the strict invariance of any actual features at the fundamental level. We can contrast this account with the familiar one provided by~\citep{lewis2}, according to which the relevant counterfactual world resembles ours up to a short period before the counterfactual event, and has slightly \emph{different laws} of nature from ours.} 
	

This raises the question of how to formulate the principle of free recombination in a way that is compatible with hyperdeterminism.
Clearly a reasonable formulation of it should be violable, or otherwise it would be vacuous. 
We propose such a formulation here in the spirit of \citet{ES}, but tailored towards field theory.
In a field theory, one associates to every open spacetime region $U$ a set of field configurations $\mathcal{F}(U)$ (either merely kinematically conceivable configurations or already constrained by dynamical laws).
Given a subregion $V \subseteq U$, we can restrict these field configurations, which gives us a map $\mathcal{F}(U) \to \mathcal{F}(V)$.
These sets of field configurations may or may not satisfy the \emph{sheaf condition} defined as follows ~\citep{bredon}:
\begin{definition} [Sheaf condition]
Given any family of regions $(U_i)_{i \in I}$ and any collection of field configurations $(f_i \in \mathcal{F}(U_i))_{i \in I}$ such that the $f_i$ agree on all possible overlaps $U_i \cap U_j$, there is a unique field configuration in $\mathcal{F}(\cup_{i \in I} U_i)$ which restricts to the given $f_i$. 
\end{definition}
\noindent In particular if $U$ and $V$ are disjoint regions, then any two field configurations $f \in \mathcal{F}(U)$ and $g \in \mathcal{F}(V)$ can be patched together to a unique field configuration in $\mathcal{F}(U \cup V)$.\footnote{This is an instance of the sheaf condition under the additional assumption that $F(U \cap V) = F(\emptyset)$ is a singleton, which is typically satisfied: there is a unique field configuration on the empty region, namely the trivial one.}
Our proposal is that this sheaf condition makes precise the principle of free recombination in the context of field theory.
It is well-known that the sheaf condition \emph{is satisfied} for analytic functions.\footnote{See e.g.~\citep[126]{KS}. Note that this does not contradict the identity theorem: given disjoint regions $U$ and $V$, \Cref{identity_theorem} does not apply since $U \cup V$ is not connected (and we are not trying to extend $f$ and $g$ to all of spacetime). If $U \cap V$ do have nonempty overlap, then \Cref{identity_theorem} may apply, but now $f \in \mathcal{F}(U)$ and $g \in \mathcal{F}(V)$ can be patched together as per the sheaf condition only if they agree on $U \cap V$.}
Therefore the principle of free recombination in this sense is indeed compatible with hyperdeterminism.
Moreover, if we try to strengthen the sheaf condition to rule out analytic functions, we will likely rule out other `innocuous' classes of functions as well.\footnote{
For example, we might require that every field configuration on a subregion $V \subseteq U$ must be extendable to all of $U$ (the restriction maps $\mathcal{F}(U) \to \mathcal{F}(V)$ must be surjective). This requirement indeed fails for analytic functions, but it fails just as well for continuous functions and for smooth functions.}


A possible third objection against hyperdeterminism---or rather a discomfort with it---may be the sentiment that it leads to non-locality. But upon a closer inspection, this sentiment is clearly distinguished from the usually discussed types of non-locality in physics, which are considered problematic by some. For instance, 
it is clearly distinguished from Bell non-locality---whether problematic or not---
which refers to the violation of local realism in quantum physics witnessed by Bell's theorem~\citep{BCPSW}, since our analytic function physics is purely classical.
Another type of non-locality that is deemed more objectionable in physics involves (typically instantaneous) non-local interactions, like the gravitational force in Newtonian gravity. This non-locality is a dynamical feature of a physical theory (that is, distinct spacetime points appear jointly in dynamical laws), while hyperdeterminism is kinematical in nature and is not concerned with how fields interact.\footnote{One may point out a dialectical inconsistency in our arguments. Here we distinguish between dynamical and kinematic non-locality, and argue that the former is more obviously objectionable in physics. But previously we have argued (though as a minor point) that the distinction between kinematic and dynamical conditions is often blurry, citing Gauss's law as an example. Yet these two claims are not strictly contradictary and we uphold that the dynamical non-locality in the gravitational force law cannot be reformulated as a kinematic one.} For example, classical electrodynamics has all interactions propagating locally through spacetime at the speed of light, even if we require all field configurations to be analytic. Mathematically, this locality manifests itself in the fact that the field equations are partial differential equations, and in particular they do not impose any direct constraints on field values at distant spacetime points. So, dynamical non-locality does not account for what is intuitively problematic about kinematic non-locality, which we conjecture comes from the apparent failure of free recombination---which is already discussed. 

Perhaps it is appropriate to conclude this section with a quote from Leibniz's \emph{Monadology}~\citep{leibniz}, where he advocated a metaphysics that involves \emph{monads}.
These are simple substances of the universe, each encoding the whole universe in pre-determined harmony with each other:

\begin{quote}
	Now this connexion or adaptation of all created things to each and of each to all, means that each simple substance has relations which express all the others, and, consequently, that it is a perpetual living \emph{mirror of the universe}. (\S 56; italic ours)
\end{quote}

\section{Conclusion}

Should general relativity be formulated in terms of analytic functions instead of smooth ones to avoid indeterminism?
Perhaps.
However, while we have argued that analytic functions are not obviously objectionable, there are no decisive reasons in their favor either.
Our intention is not to adjudicate between analytic vs.~smooth functions, but to raise (further) awareness of (1) the plenitude of options in mathematical formalism,  with the analytic framework as an example, and (2)  how an even seemingly insignificant part of the formalism of a physical theory can have significant ramifications on its philosophical outlook, with the hole argument as a cautionary tale. 
Questions like whether our world is deterministic are certainly interesting to ponder in their own right, but it would be wrong to think them as decisively informed by our physical theories. Of course, many cautionary tales of this kind have been told. But it is still useful to dislodge the particular default belief that smooth functions constitute a privileged class of functions most amenable to contemporary practice of physicists, and to counter undue realistic significance attached to them.\footnote{This then echoes the thesis of \citep{manchak}, according to which we should be \textit{pluralists} about the privileged class of models for general relativity, and in general the spirit of naturalized metaphysics discussed in \citep{ney} and \citep{esfeld}, according to which we can benefit from ``(a) elucidating the conceptual entailments of those commitments that result from what is indispensable to physical science and (b) filling in the ontologies arrived at via indispensability projects''~\citep[p.~69]{ney}.}
It is beneficial to inquire, for each piece of theoretical apparatus of our best physical theories, whether it is indeed indispensable to the theory, cautioning us against undue metaphysical commitments.


\printbibliography

\end{document}